\documentclass[aps,11pt,preprintnumbers,nofootinbib,floatfix]{revtex4}

\usepackage{graphicx}
\usepackage{epsfig}
\usepackage{dcolumn}
\usepackage{subfigure}
\usepackage{multirow}
\usepackage{amsmath}
\usepackage{amssymb}

\def\be{\begin{equation}}
\def\ee{\end{equation}}
\def\bea{\begin{eqnarray}}
\def\eea{\end{eqnarray}}

\begin{document}

\title{Topology in thermodynamics of regular black strings with Kaluza-Klein reduction}

\author{Tran N. Hung}
\email{hung.tranngoc@phenikaa-uni.edu.vn}  
\affiliation{Phenikaa Institute for Advanced Study and Faculty of Fundamental Sciences, Phenikaa University, Yen Nghia, Ha Dong, Hanoi 12116, Vietnam}
\author{Cao H. Nam}
\email{nam.caohoang@phenikaa-uni.edu.vn}  
\affiliation{Phenikaa Institute for Advanced Study and Faculty of Fundamental Sciences, Phenikaa University, Yen Nghia, Ha Dong, Hanoi 12116, Vietnam}
\date{\today}

\begin{abstract}
We study the topological defects in the thermodynamics of regular black strings (from a four-dimensional perspective) that is symmetric under the double Wick rotation and constructed in the high-dimensional spacetime with an extra dimension compactified on a circle. We observe that the thermodynamic phases of regular black strings can be topologically classified by the positive and negative winding numbers (at the defects) which correspond to the thermodynamically stable and unstable branches. This topological classification implies a phase transition due to the decay of a thermodynamically unstable regular black string to another which is thermodynamically stable. We confirm these topological properties of the thermodynamics of regular black strings by investigating their free energy, heat capacity, and Ruppeiner scalar curvature of the state space. The Ruppeiner scalar curvature of regular black strings is found to be always negative, implying that the interactions among the microstructures of regular black strings are only attractive.

\end{abstract}

\maketitle

\section{Introduction}
Black hole thermodynamics has been extensively investigated in recent decades because it provides a bridge between general relativity (GR) and quantum gravity as well as exhibits interesting phenomena in analogy to thermodynamics. Hawking showed that like the blackbody, black holes emit thermal radiation and thus it possesses a temperature $T$ that is determined by the surface gravitation at their event horizon \cite{Hawk1975}. In addition, black holes have an entropy $S$ that is the conjugate variable of the temperature and is proportional to the horizon area \cite{Bekenstein1972}. The thermodynamic behavior of black holes is governed by the four laws \cite{Bardeen1973} which are analogous to the conventional thermodynamic laws. The thermodynamic quantities of black holes are related to each other through the Smarr relation as follows \cite{Hawk1976}
\begin{eqnarray}
M=2(TS+\Omega J)+\Phi Q,
\end{eqnarray}
where $M$ is the Arnowitt-Deser-Misner (ADM) mass of black holes treated as the internal energy, $\Omega$ is the angular velocity, $J$ is the angular momentum, $Q$ is the electric charge of black holes, and $\Phi$ is the chemical potential. Recently, the cosmological constant $\Lambda$ has been considered as a thermodynamic variable associated with the notion of pressure whose conjugate variable is the thermodynamic volume \cite{SWang2006,Kastor2009,Kastor2010,Dolan2011}. This has led to a new subfield of black hole thermodynamics known as black hole chemistry \cite{Kubiznak2017} where the ADM mass of black holes is reinterpreted as the chemical enthalpy. In the extended phase space, the black hole thermodynamics exhibits abundantly phenomena which are the $P-V$ criticality corresponding to the Van der Waals-like phase transition \cite{Kubiznak2012,Hendi2013,Cai2013,XMo2014,Hendi2016,Nam2018b,Nam2019}, the black hole heat engine \cite{Johnson2014,Setare2015,Bhamidipati2017,XMo2018,Santos2018,Zhang2019,Nam2021a}, and the Joule-Thomson expansion \cite{Aydner2017a,Aydner2017b,Mo-Li2018,QLan2018,Nam2020}.

According to GR, there exists an unphysical curvature singularity inside the horizon of black holes \cite{Hawking1973}. The presence of this singularity implies the breakdown of GR at very short distances and would be absent in an ultraviolet complete theory of quantum gravity that is unfortunately lacking. Therefore, almost resolutions for the problem of the black hole singularity proposed are based on exceptional classical sources such as the non-linear electrodynamics which have received considerable attention in the literature \cite{Ayon-Beato98,Ayon-Beato2000,Bronnikov2001,Berej2006,Gonzalez2009,Toshmatov2014,Ghosh2015,Hendi2015,Dehghani2017,Rincon2107,Nam2018a,Nam2018c,Singh2018,Hyun2019}.
However, in order for the source of the non-linear electrodynamics to remove the unphysical curvature singularity, it requires an unnatural fine-tuning in the parameters of Lagrangian of the non-linear electrodynamics. This makes the non-linear electrodynamics less natural.

It was recently indicated that the compactified extra dimensions could provide a way to remove the unphysical curvature singularity of black holes \cite{Ibrah2021} and understand a statistical origin of the realistic black hole entropy \cite{Nam2023}. Indeed, the compactified extra dimensions are one of the important ingredients for constructing quantum gravity in superstring/M theory and open a new window on a geometric unification of gravity and dark matter \cite{Nam2023c}, the hierarchy problem \cite{Arkani-Hamed1998,Randall1999,Nam2021}, the radiative stability of tiny cosmological constant \cite{Nam2023b}, and evidence for the AdS distance conjecture (realized in string theory) from bottom-up physics \cite{Nam2023a}. Hence, they could provide insights into quantum gravity as well as the resolutions for the issues of black hole physics. By imposing the vacuum solutions of the system of Einstein gravity coupled to the Maxwell field in five dimensions being symmetric under the double Wick rotation \cite{Miyamoto2006,Stotyn2011}, the authors in Ref. \cite{Ibrah2021} have found a regular black string that consists of the event horizon but not the curvature singularity. The absence of the curvature singularity here is due to the presence of a bubble behind the horizon where a spacelike Killing vector shrinks at the radius of the bubble and hence it leads to the end of the spacetime there. It was also indicated the evaporation of regular black string is consistent with the unitarity principle \cite{Hung2023}.

In the free energy landscape, black holes are considered as the thermodynamic states of a canonical ensemble \cite{RanLi2020}. The generalized free energy of the system, which is denoted by $\mathcal{F}=M-S/\tau$ with the ADM mass $M$ and inverse temperature $\tau$ being two independent variables, is a function of the event horizon radius treated as the order parameter interpreting the degree of freedom of the system \cite{York1986}. The local (global) extremal points of the generalized free energy represent the on-shell states satisfying Einstein field equations corresponding exactly to the black hole solutions. The remaining states, which are not the solutions of Einstein field equations, are called the off-shell states. An on-shell state at a local (global) minimal point expresses a stable black hole, otherwise an off-shell state at a local (global) maximal point defines an unstable black hole \cite{Andr2020}. Recently, the free energy landscape has also been employed to analyze the phase transition of black holes, which is one of the most exciting subjects in the study of black hole thermodynamics. In \cite{RanLi2020}, the free energy landscape is used to study the kinetic switching process of black hole states and the Hawking-Page phase transition. This work showed that it is possible to lead the phase transition between the AdS black hole and thermal AdS. Similar results for Reissner-Nordström-Anti de Sitter (RN-AdS) black holes were also found in Ref. \cite{RanLi22020}.

Because a black hole is a solution of Einstein field equations, we can consider it as a zero point of the tensor field $\mathcal{E}_{\mu\nu}\equiv G_{\mu\nu}-\frac{8\pi G}{c^{4}}T_{\mu\nu}$ where $G_{\mu\nu}$ and $T_{\mu\nu}$ are the Einstein tensor and the energy-momentum tensor of the matter, respectively. This idea evokes that the black hole can be assigned to a topological charge. As a result, we can determine the local and global characteristics of the black hole without knowing its detailed configuration. Recently, the topological charge has been introduced to study black hole thermodynamics. Each black hole is endowed with a topological charge by using the vector field $\phi$ defined as follows \cite{SWWei2022}
\begin{eqnarray}
\phi=\left(\frac{\partial \mathcal{F}}{\partial r_{h}},-\cot \Theta \csc \Theta \right) ,
\label{eqn:phi}
\end{eqnarray}
where $r_h$ is the radius of the event horizon and the parameter $\Theta$ ($0\leq\Theta\leq\pi$) is introduced by the reason of the axis limit \cite{Cunha2020}. It should be noted that the vector field $\phi$ would point outward at the points with $\Theta=0$ and $\pi$ due to the divergence of the component $\phi^\Theta$ at these points. Using Duan's $\phi$-mapping topological current \cite{Duan1979,LBFu2000}, we construct a conserved topological current $j^{\mu}$ satisfying $\partial_{\mu}j^{\mu}=0$ and given by
\begin{eqnarray}
 j^\mu=\frac{1}{2\pi}\epsilon^{\mu\nu\rho}\epsilon_{ab}\frac{\partial n^a}{\partial y^\nu}\frac{\partial n^b}{\partial y^\rho},   
\end{eqnarray}
where $\epsilon^{\mu\nu\rho}$ and $\epsilon_{ab}$ are the totally anti-symmetric tensors, $y^\nu=(\tau,r_h,\Theta)$, and the unit vector $n^a$ is defined by $n^a=\phi^a/||\phi||$ with $\phi^a=(\phi^{r_h},\phi^\Theta)$ and $||\phi||$ denoted the norm of the vector $\phi$. This topological current is only nonzero at the zero points where the vector field $\phi$ vanishes. Interestingly, the zero points of the vector field $\phi$ correspond to $\Theta=\pi/2$ and $\tau=T^{-1}$, where $T$ is the black hole temperature. This means that each zero point would represent a black hole solution. For a $i$-th zero point $z_i$ of the vector field $\phi$, we can calculate the winding number $\omega_{i}=\beta_i\eta_i$ where $\beta_i$ counts the number of loops around $z_i$ and $\eta_i=\text{sign}(J^0|_{z_i})=\pm1$ is the Brouwer degree with 
$j^\mu$ to be the Jacobi tensor related to the topological current as $j^\mu=\delta^2(\phi)J^\mu(\phi/y)$. The winding numbers $w_i$ reveal the local topological properties which are related to the aspect of thermodynamic stability: the positive/negative winding numbers correspond to the stable/unstable black holes \cite{SWWei2022}. These winding numbers can also be computed by determining the sign of the residues of a characterized complex function $\mathcal{R}(z)$ which is constructed from the generalized free energy of the system \cite{FangZhang2023}. The global topological winding number $W$ is derived by summing the winding numbers of all black holes in the different branches as $W=\int_\Sigma j^0d^2y=\sum^N_{i=1}w_i$ where $\Sigma$ refers to the loop containing the entire parameter space. Studying thermodynamic topological properties of the Schwarzschild, RN, and RN-AdS black holes, it was pointed out that the global topological number $W$ can take three values -1, 0, and 1.

The topological approach outlined above has recently received attention due to its simplicity in investigating the thermodynamic features of black holes. This approach was used to study the Hawking-Page phase transition of the Schwarzschild-AdS black hole and its holographic dual corresponding to the confinement-deconfinement transition \cite{Yerra2022}. Taking into account the quantum gravity corrections expressed by the presence of the higher-derivative terms, the topological classification of the thermodynamic phases and the corresponding phase transitions have been analyzed for black holes in Einstein-Gauss-Bonnet gravity \cite{CLiu2023,Liwang2023} and Lovelock gravity \cite{NCBai2023}. However, these cases are all static black holes. With respect to more realistic black holes or astronomical black holes, the topological approach was employed to study the thermodynamics of the rotating black holes \cite{DiWu2023,DiWu2023-2}. The results in these papers have supported the conjecture proposed by Ref. \cite{SWWei2022}.

This paper is organized as follows. In Sect. \ref{secreg}, we briefly review the solution of the regular black string in the five-dimensional Einstein-Maxwell theory and its dimensional reduction on a circle. In Sect. \ref{sec:ther}, we investigate the topological thermodynamics of the regular black string from the external four-dimensional perspective corresponding to the Kaluza-Klein reduction along the compact fifth dimension. We use the Ruppeiner geometry approach in order to confirm the results obtained in Sect. \ref{sec:thergeo} based on the generalized free energy. In Sect. \ref{sec:highdim}, we study the effects of the dimensional number on the topological behavior of the thermodynamics of the dimensionally reduced regular black string. Finally, we make conclusions in Sect. \ref{sec:conclu}.

\section{\label{secreg} Regular black string solution}
\begin{figure}[h]
  \includegraphics[scale=0.2]{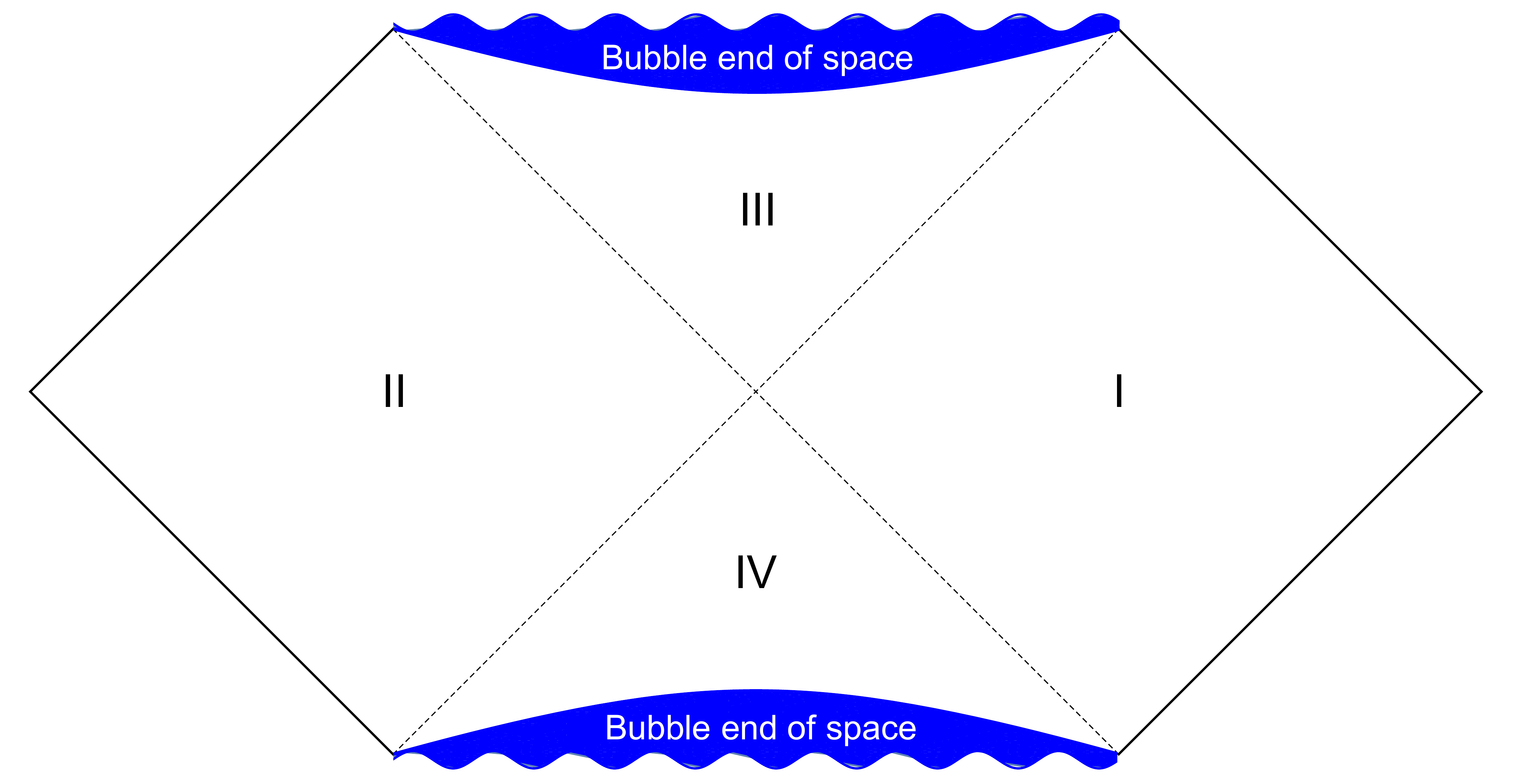}
  \caption{The Penrose diagram of the regular black string.}
  \label{figPendiag}
\end{figure}

In this section, we review the regular black string solution of the five-dimensional Einstein-Maxwell theory compactified on a circle $S^1$ \cite{Ibrah2021}. The action of this system is given by
\begin{eqnarray}
S_{5D}=\int d^5X\sqrt{-g_5}\left(\frac{1}{2\kappa_{5}^{2}}R_5-\frac{1}{4}F_{MN}F^{MN}\right),
\end{eqnarray}
where $\kappa_{5}$ is the five-dimensional gravitational coupling, $R_5$ is the Ricci scalar, and $F_{MN}=\partial_M A_N-\partial_N A_M$ is the field strength tensor of the Maxwell field $A_M$. By varying the above action, we find the equations of motion as follows
\begin{eqnarray}
R_{MN}-\frac{1}{2}R_5g_{MN}&=&\kappa^2_5\left[F_{MP}{F_N}^P-\frac{g_{MN}}{4}F_{PQ}F^{PQ}\right],\\
\nabla_N F^{MN}&=&0.
\end{eqnarray}
We find a solution of magnetic charge which is spherically symmetric and described by the following ansatz
\begin{eqnarray}
ds^{2}_5&=&-f_{S}(r)dt^{2}+f_{B}(r)dy^{2}+\frac{dr^{2}}{h(r)}+r^{2}\left(d\theta^{2}+\sin^{2}\theta d\phi^{2}\right),\nonumber\\
F&=&P\sin\theta d\theta\wedge d\phi,\label{WDR-sol}
\end{eqnarray}
where $y$ is the fifth compactified dimension that is periodic with a period $2\pi R_y$ and $P$ is a constant that is the magnetic charge of the system. By imposing the double Wick rotation \cite{Miyamoto2006,Stotyn2011} given by $(t,y,r_h,r_b)\rightarrow(it,it,r_b,r_h)$ (where the timeline Killing vector $\partial_t$ and spacelike Killing vector $\partial_y$ shrink at $r_h$ and $r_b$, respective) to be a symmetry of the solution, we find
\begin{eqnarray}
f_{B}(r)=1-\frac{r_{b}}{r},\quad f_{S}(r)=1-\frac{r_{h}}{r},\\
h(r)=f_{B}(r)f_{S}(r),\quad P=\pm\frac{1}{\kappa_{5}}\sqrt{\frac{3r_{b}r_{h}}{2}}.\label{exp-fSB}
\end{eqnarray}
The solution has two coordinate singularities at $r_h$ and $r_b$. In the case o $r_h>r_b$, the coordinate singularity $r=r_h$ corresponds to an event horizon, whereas the coordinate singularity $r=r_b$ that is hidden behind the event horizon corresponds to the end of spacetime as a smooth bubble $S^2$ sitting at the apex of the cone in the two-dimensional Minkowski space $\mathbb{R}^{1,1}$. This means that the black string solution would not have the region of the conventional curvature singularity and as a result, the solution leads to a regular black string. This can easily be seen in the causal structure of spacetime corresponding to the geometry of the regular black string as depicted in Fig. \ref{figPendiag}.

In order to find the temperature and the Bekenstein-Hawking entropy of the regular black string, we study the near-horizon geometry in the Euclidean signature (obtained by the Wick rotation as $t=it_E$) which is given by
\begin{eqnarray}
ds^2=\frac{r_h-r_b}{4r^3_h}\rho^2dt^2_E+d\rho^2+\frac{r_h-r_b}{r_h}dy^2+r^2_h\left(d\theta^{2}+\sin^{2}\theta d\phi^{2}\right),
\end{eqnarray}
where $\rho\equiv2\left[(r-r_h)/(r_h-r_b)\right]^{1/2}r_h\rightarrow0$. The regularity of the near-horizon geometry at the origin (corresponding to $r=r_h$) implies the period as $t_E\rightarrow t_E+\beta$ where $\beta=1/T$ is the inverse temperature of the regular black string and given by
\begin{eqnarray}
T=\frac{1}{4\pi r_{h}}\sqrt{1-\frac{r_{b}}{r_{h}}}.
\end{eqnarray}
In addition, from the near-horizon geometry, it is easy to find a topology of the event horizon as $ S^{1}\times S^{2} $ whose radii are $\sqrt{1-r_h/r_b}R_y$ and $r_h$, respectively. Hence, we can compute the Bekenstein-Hawking entropy of the regular black string as
\begin{eqnarray}
S=\frac{8\pi^{2}}{\kappa_{4}^{2}}\sqrt{r_{h}^{3}(r_{h}-r_{b})}.
\end{eqnarray}

Under the dimensional reduction on the circle, the black string is the solution of equations of motion corresponding to the following action of the Einstein-Maxwell-scalar theory
\begin{eqnarray}
S_{4D}=\int d^4x\sqrt{-g_4}\left(\frac{1}{2\kappa_{4}^{2}}R_4-\frac{3}{\kappa^2_4}\partial_\mu\Phi\partial^\mu\Phi-\frac{e^{-2\Phi}}{4e^2}F_{\mu\nu}F^{\mu\nu}\right),
\end{eqnarray}
where $\kappa^2_4=\kappa^2_5/(2\pi R_y)$ and $e^2=1/(2\pi R_y)$. This identification implies the expressions for the four-dimensional metric, Maxwell field, and dilaton as follows
\begin{eqnarray}
e^{2\Phi}&=&f^{1/2}_B,\nonumber\\
F&=&\pm\frac{e}{\kappa_4}\sqrt{\frac{3r_{b}r_{h}}{2}}\sin\theta d\theta\wedge d\phi,\\
ds^2_{4}&=&f^{1/2}_B\left(-f_Sdt^2+\frac{dr^2}{f_Bf_S}+r^2d\Omega^2\right).\nonumber
\end{eqnarray}
Then, from the four-dimensional viewpoint, we can determine the mass $M$ and the magnetic charge $Q$ of the black string as
\begin{eqnarray}
M=2\pi\left(\frac{2r_{h}+r_{b}}{\kappa_{4}^{2}} \right) ,\quad Q=\sqrt{\frac{3r_{b}r_{h}}{2\kappa_{4}^{2}}}.
\end{eqnarray}
We can rewrite $M$, $S$, and $T$ as the functions of the horizon radius $r_{h}$ and the magnetic charge $Q$ as follows
\begin{eqnarray}
M=4\pi \left(r_{h}+\frac{Q^{2}}{3r_{h}} \right) ,\label{ADMmass}\\
S=8 \pi^{2}r_{h}\sqrt{rh^{2}-\frac{2Q^{2}}{3}},\label{BHentropy}\\
T=\frac{1}{4\pi r_{h}}\sqrt{1-\frac{16\pi}{3}\frac{Q^{2}}{r_{h}^{2}}}.
\end{eqnarray}
For the sake of simplicity, we rescale the length dimension by substituting $\kappa_{4}=1$ in the above equations. These thermodynamic quantities satisfy the first law of black string thermodynamics as follows
\begin{eqnarray}
\delta M=T\delta S+\Phi \delta Q,
\label{eqn:fl}
\end{eqnarray}
where $\Phi$ is the magnetic potential and $M$ can be interpreted as the internal energy of the system. 

\section{Topological thermodynamics}
\label{sec:ther}
In order to study the topological thermodynamics of the regular black string from the four-dimensional viewpoint, we first need to introduce the generalized free energy given as
\begin{eqnarray}
\mathcal{F}&=&M-\frac{S}{\tau}\nonumber\\
&=&4\pi r_{h}+\frac{4\pi Q^{2}}{3r_{h}}-\frac{8\pi^{2} r_{h}}{3\tau}\sqrt{9r_{h}^{2}-6Q^{2}},
\end{eqnarray}
where the ADM mass $M$ and the Bekenstein-Hawking entropy $S$ of the regular black string are given in Eqs. (\ref{ADMmass}) and (\ref{BHentropy}), respectively. Note that, if the parameter $\tau$ is equal to the inverse of the Hawking temperature of the regular black string, then the generalized free energy is on-shell.

From the generalized free energy and following Eq. (\ref{eqn:phi}), we construct a vector field $\phi$ whose components read
\begin{eqnarray}
\phi^{r_{h}}&\equiv &\frac{\partial \mathcal{F}}{\partial r_{h}} \nonumber \\
&=&4\pi(Q^{2}-3r_{h}^{2})\left(\frac{4\sqrt{3}\pi}{3\tau\sqrt{3r_{h}^{2}-2Q^{2}}}-\frac{1}{3r_{h}} \right),\\
\phi^\Theta&=&-\cot \Theta \csc \Theta.
\end{eqnarray}
We observe that the zero points of the vector field $(\phi^{r_{h}},\phi^{\Theta})=(0,0)$ would correspond to the constraint $\Theta=\pi/2$ and $\tau=T^{-1}$ which represents the on-shell regular black string. 
\begin{figure}[h]
  \includegraphics[scale=0.6]{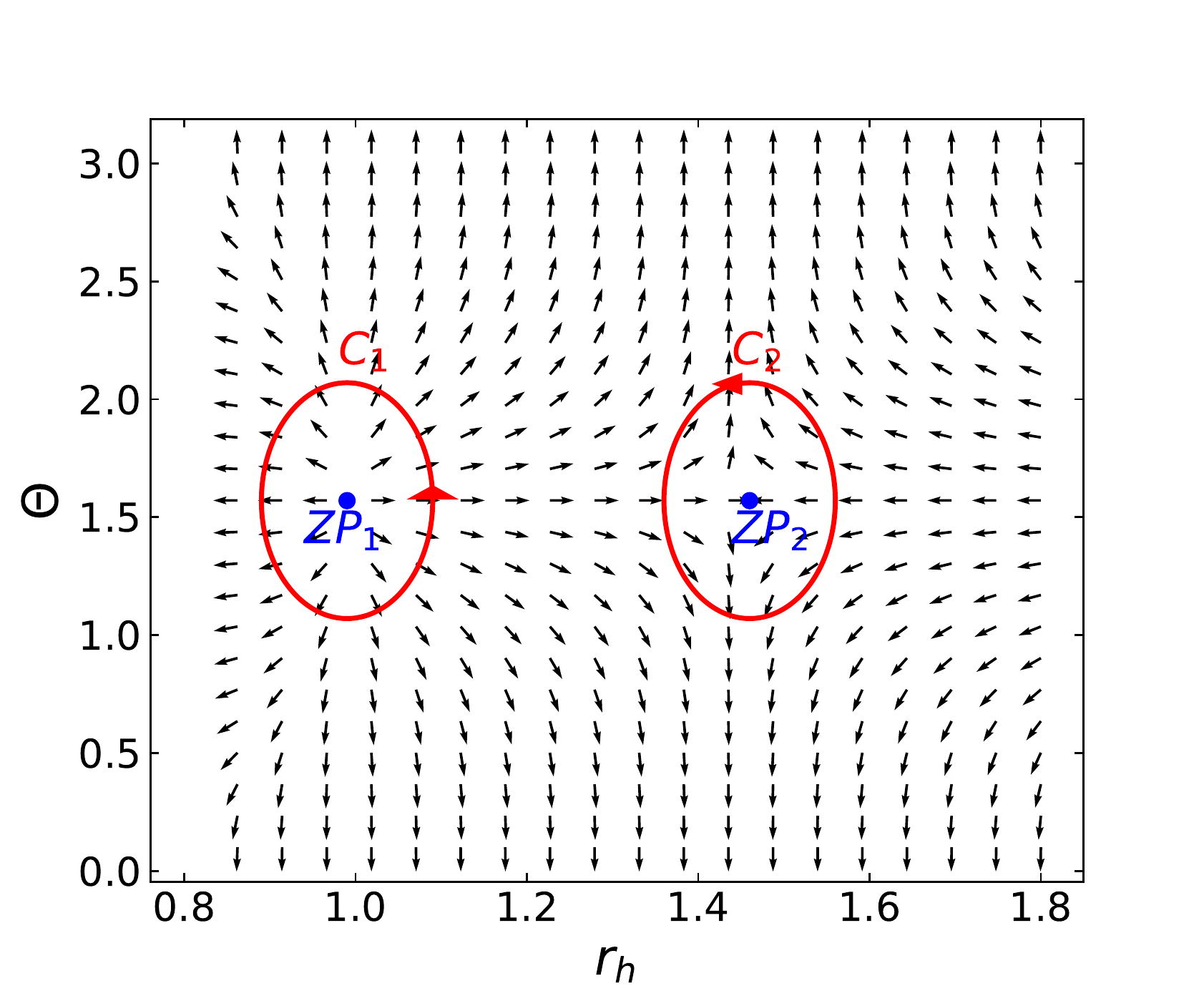}
  \caption{The black arrows represent the unit vector field on a $r_{h}-\Theta$ plane. There are two zero points $ZP_{1}$ and $ZP_{2}$ at the coordinates $(0.99,\frac{\pi}{2})$ and $(1.44,\frac{\pi}{2})$, respectively. Two red closed loops $C_{1}$ and $C_{2}$ enclose $ZP_{1}$ and $ZP_{2}$, respectively.}
  \label{figf1}
\end{figure}
In Fig. \ref{figf1}, we depict the unit vector $n^a=(\phi^{r_h}/||\phi||,\phi^\Theta/||\phi||)$  with $Q=1$ and $\tau=7\pi$, where the norm of the vector $\phi$ is defined as
\begin{eqnarray}
    ||\phi||=\sqrt{(\phi^{r_h})^2+(\phi^\Theta)^2}.
\end{eqnarray}
We observe that there are two zero points $ZP_{1}$ and $ZP_{2}$ located at the coordinates $(0.99,\frac{\pi}{2})$ and $(1.44,\frac{\pi}{2})$, respectively. By calculating the Brouwer degree $\text{sign}(J^{0}(\phi/y))$, we can determine the winding numbers $\omega $ for these zero points. According to the definition of the vector field $\phi$, we evaluate the Brouwer degree as follows
\begin{eqnarray}
    \text{sign}(J^0|_{z_i})&=&\text{sign}\left( \partial_{r_h}\phi^{r_h}\partial_{\Theta}\phi^{\Theta}|_{z_i}\right) \nonumber \\
    &=&\text{sign}\left[\frac{8\pi}{3r_{h}^{3}}\left(1-\frac{18\sqrt{3}\pi r_{h}^{4}(r_{h}^{2}-1)}{(3r_{h}^{2}-2)^{3/2}\tau}\right)\csc \Theta (\cot^{2}\Theta+\csc^{2}\Theta)|_{z_i}\right].
\end{eqnarray}
Then, it is easy to find $ \omega_{1}=1 $ and $ \omega_{2}=-1 $ for $ ZP_{1} $ and $ ZP_{2} $, respectively. The global topological number of the dimensionally reduced regular black string thus is $W=0$ which is similar to the RN black hole \cite{SWWei2022}. 

\begin{figure}[h]
  \includegraphics[scale=0.4]{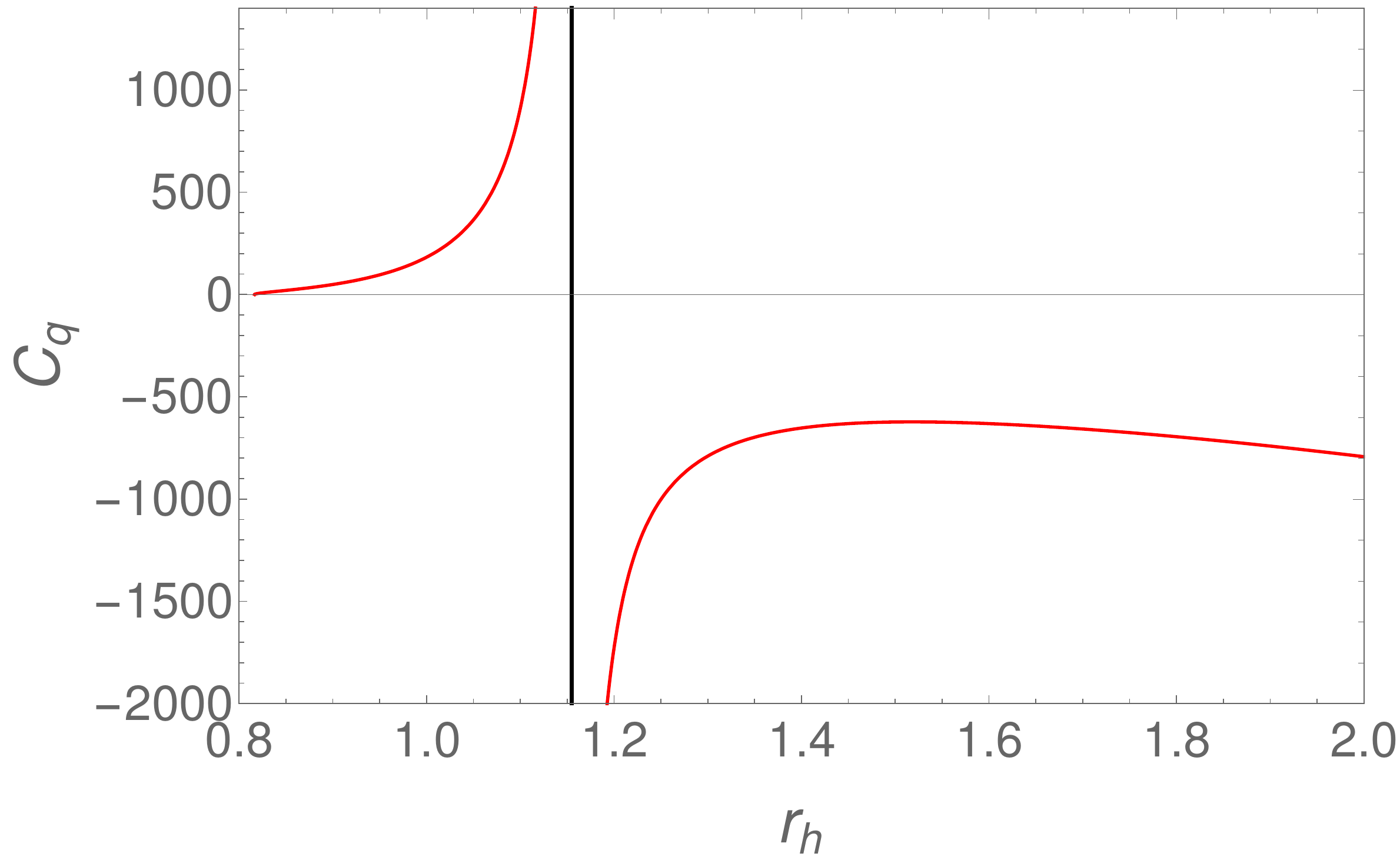}
  \caption{The heat capacity $C_{Q}$ as a function of $r_{h}$ which is positive if $r_{h}<\sqrt{4/3}$ and negative if $r_{h}>\sqrt{4/3}$. The vertical line represents the location $r_{h}=\sqrt{4/3}$ corresponding to a second-order phase transition.}
  \label{figf2}
\end{figure}

The zero points $ZP_{1}$ and $ZP_{2}$ would correspond to the stable and unstable black holes, respectively, as conjectured in \cite{SWWei2022}. In order to confirm this implication of winding numbers on the stability of the dimensionally reduced regular black string, we study the behavior of the heat capacity evaluated at a fixed charge $Q$. The heat capacity $C_Q$ can indicate the thermodynamic stability and the potential occurrence of second-order phase transitions in the thermodynamic system. More specifically, the positive heat capacity corresponds to a stable black string, whereas the negative heat capacity signifies an unstable black string. In addition, the divergences of the heat capacity imply the second-order phase transitions. In our case, the heat capacity of the dimensionally reduced regular black string at a fixed charge $Q$ is given as follows
\begin{eqnarray}
C_{Q}&\equiv &T\left(\frac{\partial S}{\partial T} \right)_{Q} =\left(\frac{\partial M}{\partial r_{h}} \right)_{Q}\left(\frac{\partial T}{\partial r_{h}} \right)_{Q}^{-1}\nonumber\\
&=&\frac{16\pi^{2}r_{h}^{2}\left(3r_{h}^{2}-Q^{2} \right) }{4Q^{2}-3r_{h}^{2}}\sqrt{1-\frac{2Q^{2}}{3r_{h}^{2}}}.
\label{eqn:cq}
\end{eqnarray}
One finds that the heat capacities $C_{Q}$ at the zero points $ ZP_{1} $ and $ ZP_{2} $ which correspond to the positive and negative winding numbers are approximately 160.26 and -634.08, respectively. The properties of the heat capacity lead us to infer that the positive (negative) winding number represents a stable (unstable) regular black string. In particular, we observe that the heat capacity $C_Q$ diverges at the critical horizon radius $r_c=\sqrt{4/3}$, which indicates the presence of a second-order phase transition point. This shows the fact that the unstable regular black string would decay to the stable one which is more thermodynamically favored. We depict the behavior of the heat capacities $C_{Q}$ in Fig. \ref{figf2} which shows that the heat capacity for the small black strings ($r_h<r_c$) is positive, whereas the large black strings ($r_h>r_c$) takes the negative values.

\begin{figure}[h]
  \includegraphics[scale=0.4]{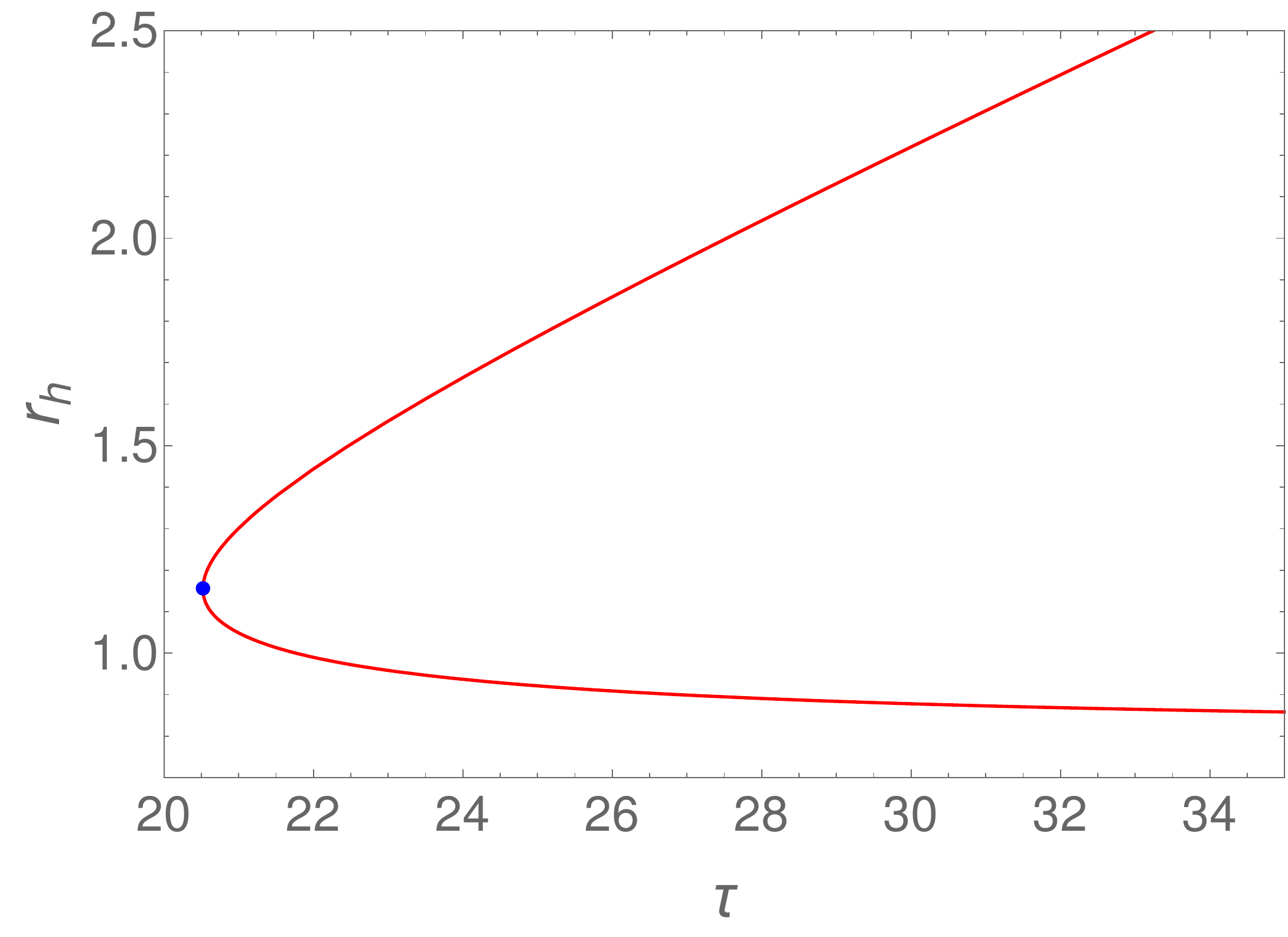}
  \caption{Locus of zero points is shown in the $r_{h}-\tau$ plane. The blue dot $(r_c,\tau_c)=(\sqrt{4/3},\sqrt{128\pi^2/3})$ divides this locus into two branches: the upper branch has the winding number -1 and the lower one has the winding number 1.}
  \label{fig:f3}
\end{figure}

The collection of zero points in the $r_{h}-\tau$ plane is represented by a red curve as illustrated in Fig. \ref{fig:f3}. 
This curve is divided into two branches by a (blue) critical point $(r_c,\tau_c)=(\sqrt{4/3},\sqrt{128\pi^2/3})$. The upper branch corresponds to the large black strings with the winding number -1, while the lower branch corresponds to the small black strings with the winding number 1. Notably, at this critical point, we find that $\frac{d^{2}\tau}{dr_{h}^{2}}=8\sqrt{6}\pi>0$, indicating that it serves as a generation point for the appearance of new phase \cite{SWWei2022}. This demonstrates that the unstable phase would undergo a thermodynamic decay to the stable phase at the second-order phase transition point which can be confirmed by the behavior of the heat capacity $C_Q$.

\begin{figure}[h]
  \includegraphics[scale=0.4]{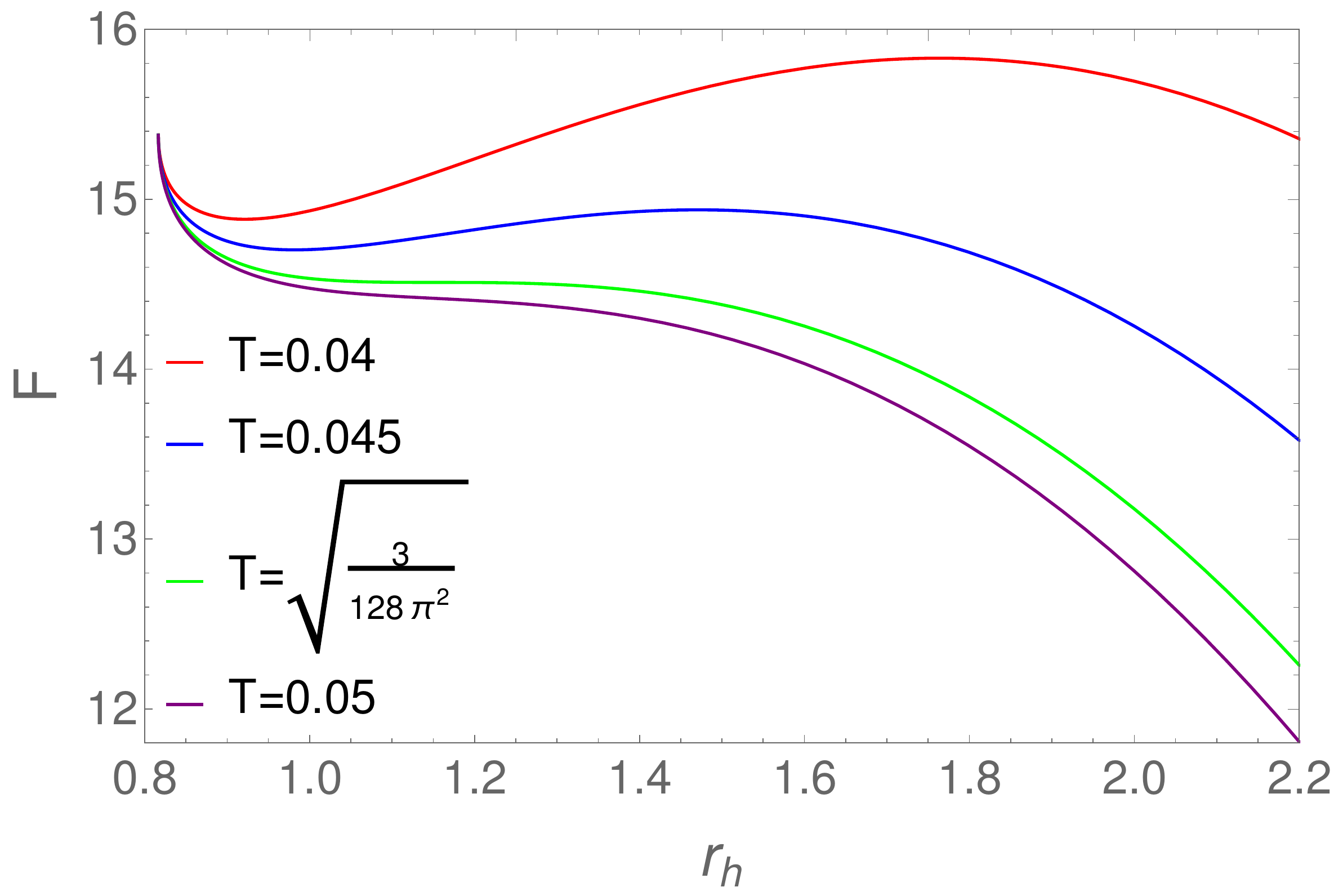}
  \caption{The generalized free energy $\mathcal{F}$ as a function of $r_{h}$ for various black string states under different values of the temperature $T=\tau^{-1}$.}
  \label{fig:f4}
\end{figure}

\begin{figure}[h]
  \includegraphics[scale=0.4]{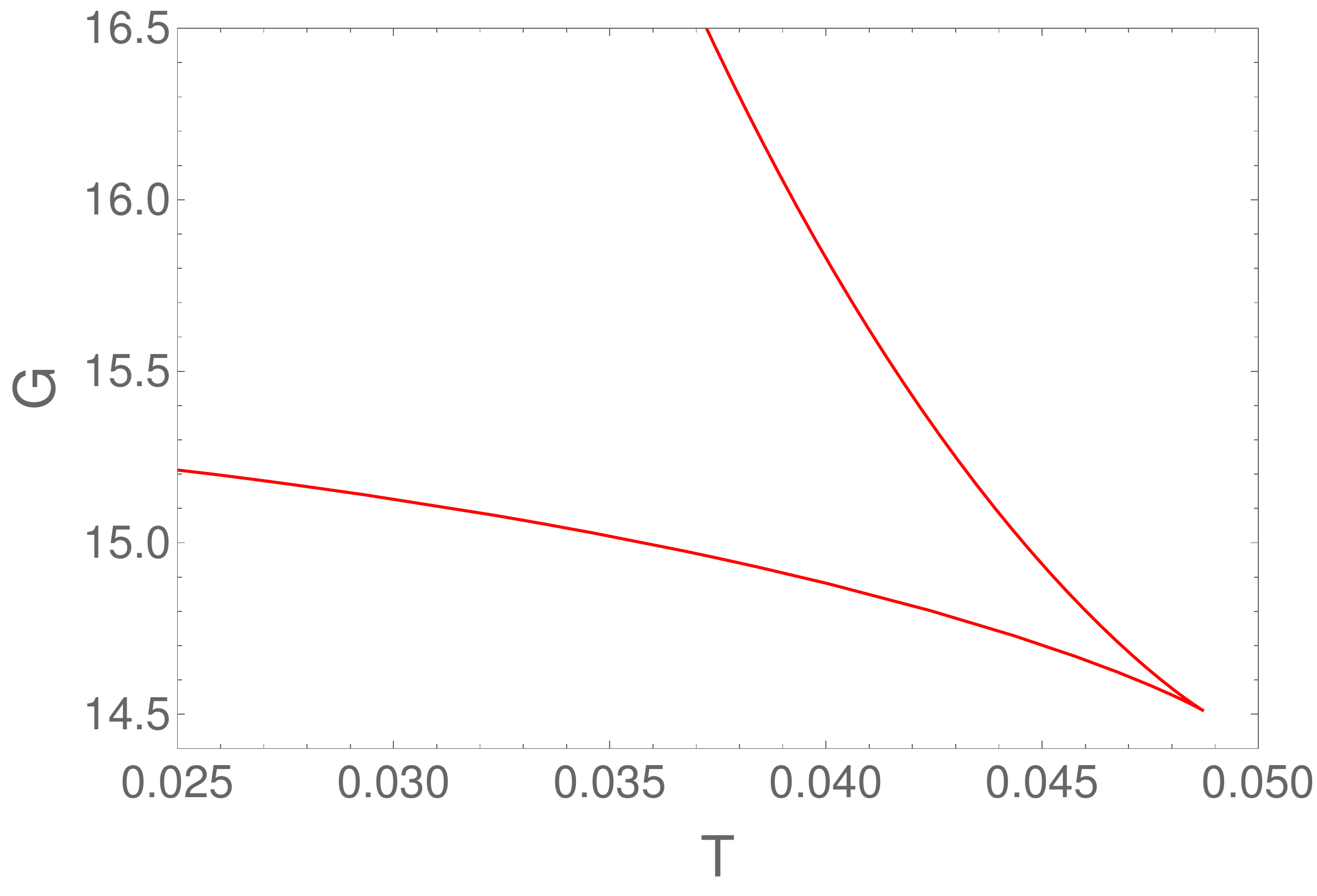}
  \caption{The on-shell free energy $G$ for the dimensionally reduced regular black string versus its temperature $T$.}
  \label{fig:f5}
\end{figure}

In order to validate the previous results, we can examine the behavior of the generalized free energy. We investigate the relationship between the generalized free energy $\mathcal{F}$ and the horizon radius $r_{h}$ at various temperature values represented by $\tau^{-1}$, as depicted in Fig. \ref{fig:f4}. 
The extremal points are obtained by setting the derivative of $\mathcal{F}$ with respect to $r_{h}$ to be equal to zero, which has an interpretation as the existence of regular black strings. Consequently, these points are determined by the following equation
\begin{eqnarray}
    -4\pi r_{h}^{2}\sqrt{9r_{h}^{2}-6}+(3r_{h}^{2}-2)\tau=0.\label{minmax}
\end{eqnarray}
 If $\tau^{-1}<\tau_{c}^{-1}$, this equation has two solutions given by
 \begin{eqnarray}
     r_{s}=\frac{\sqrt{3\tau^2-\sqrt{9\tau^4-384\pi^2\tau^2}}}{4\sqrt{6}\pi},\ \
     r_{l}=\frac{\sqrt{3\tau^2+\sqrt{9\tau^4-384\pi^2\tau^2}}}{4\sqrt{6}\pi}.
 \end{eqnarray}
which correspond to the small and large black strings, respectively. We observe that the generalized free energy exhibits a local minimum at $r_{h}=r_{s}$, indicating the presence of a stable black string. Whereas, a global maximum appears at $r_{h}=r_{l}$, corresponding to an unstable black string. As depicted in Fig. \ref{fig:f5}, the behavior of the on-shell free energy in terms of the temperature reveals a swallowtail structure that shows an abrupt change in the on-shell free energy crossing $T_c=\sqrt{3/128\pi^{2}}$. This behavior signifies the occurrence of the second-order phase transition between the small and large black strings. Additionally, if $\tau^{-1}>\tau_{c}^{-1}$, the equation (\ref{minmax}) has no solution, suggesting the absence of a black string state. We find that as $\tau^{-1}$ decreases towards the value of $\tau_c$, the two extremal points will gradually merge. As a result, the critical temperature for the phase transition point is determined by the equality of $r_{s}$ and $r_{l}$. This condition gives us the critical temperature given by $T_c=\sqrt{3/128\pi^{2}}$. 

\section{\label{sec:thergeo} Thermodynamic geometry of regular black string}

Besides the traditional thermodynamics method, we can use the Riemannian geometry on the space of states to study the thermodynamic system near critical points where  the classical fluctuation theory fails due to taking into account the local correlations. The curvature of the geometry on the space of states is connected to the interactions of the underlying constituents of the thermodynamic system. This curvature is zero for the case of the ideal thermodynamic system corresponding to the non-interaction system. In this sense, the thermodynamic fluctuations are entered in the curvature of the geometry on the space of states and hence one believes that the geometry can reveal some critical properties of the thermodynamic system. Considering the internal energy as a function of the extensive variables, Weinhold constructed a Riemannian metric endowed on the state space as follows \cite{Wein1975}
\begin{eqnarray}
g_{ij}^{W}\equiv\frac{\partial^{2} U}{\partial X^{i} \partial X^{j}},
\label{eqn:w}
\end{eqnarray}
where $U$ is the internal energy and $X^{i}$ are the extensive variables of the thermodynamic system. By using another geometrical metric in thermodynamic fluctuation theory related to the second derivatives of the entropy function, Ruppeiner introduced a new metric as follows \cite{Ruppeiner1983,Rupp1995}
\begin{eqnarray}
g_{ij}^{R}\equiv-\frac{\partial^{2} S}{\partial X^{i} \partial X^{j}},
\label{eqn:r}
\end{eqnarray}
where $S$ is the entropy function of the thermodynamic system. Although the two metrics are different, the line elements in the Weinhold and Ruppeiner geometries are conformally related by \cite{Mrug1984,Salamon1984}
\begin{eqnarray}
ds_{R}^{2}=\frac{1}{T} ds_{W}^{2},
\label{eqn:wr}
\end{eqnarray}
where $T$ is the temperature of the thermodynamic system. As a result, the equations (\ref{eqn:w}) and (\ref{eqn:wr}) are usually used to calculate the Ruppeiner metric $g_{ij}^{R}$ of black holes for the sake of simplicity.

Now let us calculate the scalar curvature of the thermodynamic space of the dimensionally reduced regular black string and find its singularities to confirm the phase transition point obtained previously. Considering the AMD mass $M$ of the dimensionally reduced regular black string as a function of the entropy $S$ and the charge $Q$, we define $\bar{M}(S,\Phi)$ as a new conjugate potential of $M(S,Q)$ by the following Legendre transformation
\begin{eqnarray}
\bar{M}(S,\Phi)=M(S,Q)-\Phi Q.
\end{eqnarray}
The first law of thermodynamics of the dimensionally reduced regular black string is written as follows
\begin{eqnarray}
d\bar{M}=TdS-Qd\Phi.
\end{eqnarray}
From this first law, we derive the Maxwell relation as
\begin{eqnarray}
\left(\frac{\partial T}{\partial \Phi} \right)_{S}=-\left( \frac{\partial Q}{\partial S}\right)_{\Phi}. 
\end{eqnarray}
In the state space defined by the thermodynamic quantities $(S,\Phi)$, the components of the Ruppeiner metric are given by
\begin{eqnarray}
\bar{g}_{SS}^{R}&=&\frac{1}{T}\frac{\partial^{2}\bar{M}}{\partial^{2}S}=\frac{1}{T}\frac{\partial}{\partial S}\left(\frac{\partial \bar{M}}{\partial S} \right)_{\Phi}=  \frac{1}{T}\left(\frac{\partial T}{\partial S} \right)_{\Phi},\\
\bar{g}_{S\Phi}^{R}=\bar{g}_{\Phi S}^{R}&=&\frac{1}{T}\frac{\partial^{2}\bar{M}}{\partial \Phi \partial S}=\frac{1}{T}\frac{\partial}{\partial \Phi}\left( \frac{\partial \bar{M}}{\partial S}\right) _{\Phi}=\frac{1}{T}\left(\frac{\partial T}{\partial \Phi} \right)_{S} =-\frac{1}{T}\left( \frac{\partial Q}{\partial S} \right) _{\Phi},\\
\bar{g}_{\Phi \Phi}^{R}&=&\frac{1}{T}\frac{\partial^{2}\bar{M}}{\partial^{2}\Phi}=\frac{1}{T}\frac{\partial}{\partial \Phi}\left(\frac{\partial \bar{M}}{\partial \Phi} \right)_{S}=   -\frac{1}{T}\left( \frac{\partial Q}{\partial \Phi}\right)_{S}.  
\end{eqnarray}
Then, we can calculate the components of the Ruppeiner metric endowed on the state space of the dimensionally reduced regular black string as follows
\begin{eqnarray}
\bar{g}_{SS}^{R}&=&-\frac{1}{2S},\\
\bar{g}_{S\Phi}^{R}&=&\bar{g}_{\Phi S}^{R}=-\frac{3\Phi}{2(24\pi^2-\Phi^{2})},\\
\bar{g}_{\Phi \Phi}^{R}&=& \frac{3S(-48\pi^2+\Phi^{2})}{2(24\pi^2-\Phi^{2})^{2}}.  
\end{eqnarray}
Accordingly, it is easy to find the scalar curvature as
\begin{eqnarray}
R=-\frac{6\pi^2(24\pi^2-\Phi^{2})}{S(12\pi^2-\Phi^{2})^{2}}.
\end{eqnarray}
Alternatively, we can express the scalar curvature as a function of the horizon radius $r_{h}$ and the charge $Q$ as
\begin{eqnarray}
R=-\frac{3\sqrt{-6 Q^{2}r_{h}^{2}+9r_{h}^{4}}}{8\pi^2\left( -4 Q^{2}+3r_{h}^{2}\right)^{2} }
\end{eqnarray}

\begin{figure}[h]
  \includegraphics[scale=0.4]{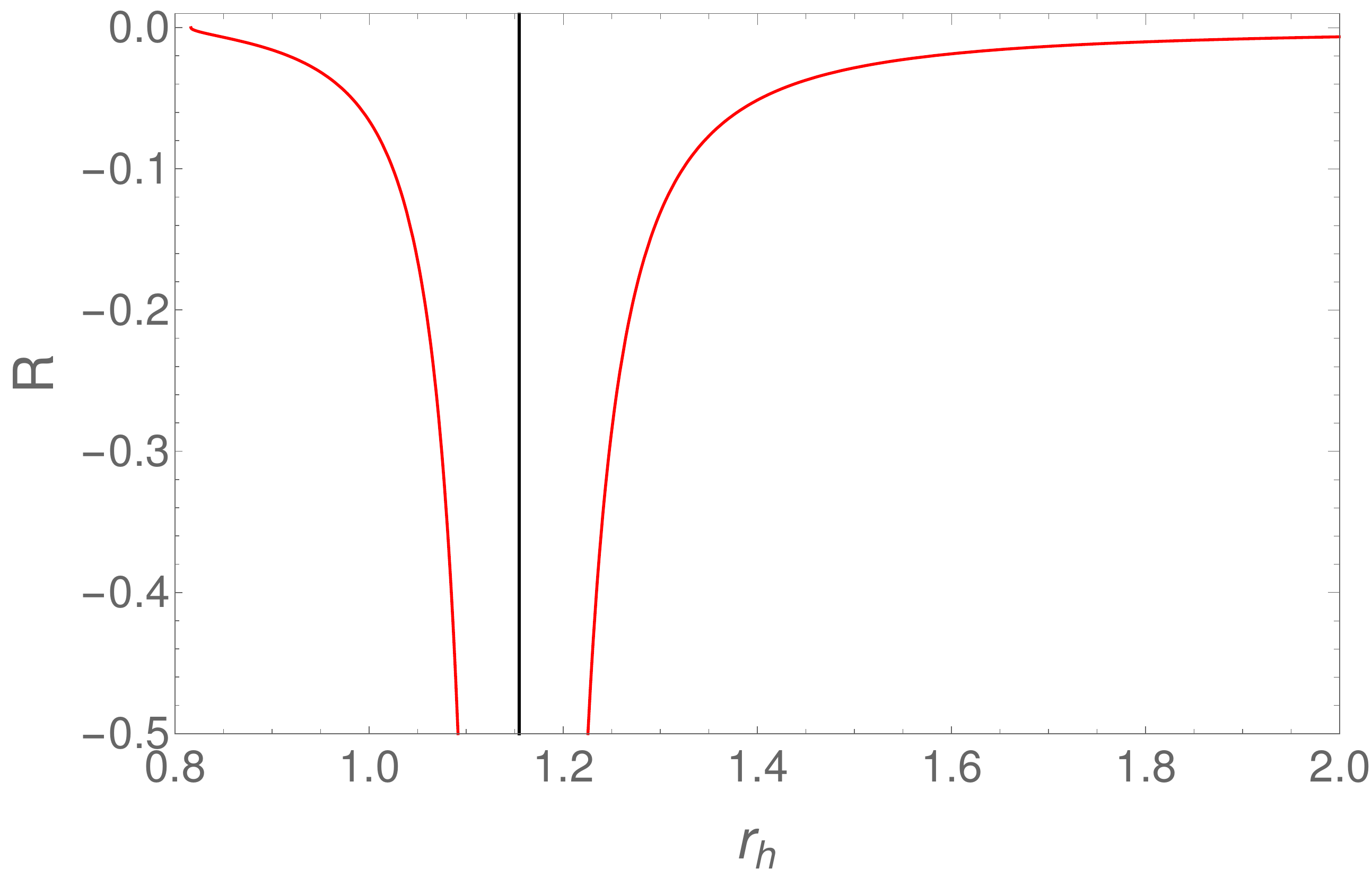}
  \caption{The Ruppeiner scalar curvature $R$ in terms of $r_{h}$. The black vertical line represents the location $r_{h}=\sqrt{4/3}$.}
  \label{fig:f6}
\end{figure}

With setting the charge $Q=1$ corresponding to the scale $r_h\rightarrow r_h/Q$ and $R\rightarrow Q^2R$, the Ruppeiner scalar curvature can be considered as a function of $r_{h}$ and its behavior is sketched in Fig. \ref{fig:f6}. 
We observe that the Ruppeiner scalar curvature diverges at $r_{c}=\sqrt{4/3}$ where the heat capacity $C_{Q}$ is too. Therefore, this result confirms the thermodynamic instability of the dimensionally reduced regular black string corresponding to the presence of the stable and unstable phases (as indicated by the investigation of the topological thermodynamics) leading to the second-order phase transition between them. 

In addition, the scalar curvature $R$ is a measure of the fluctuations in the thermodynamic system caused by the presence of the interactions. From Fig. \ref{fig:f6}, we see that the scalar curvature of the state space for the dimensionally reduced regular black string is always negative. This means that the microstructures of the dimensionally reduced regular black string should only the attractive interactions.

\section{\label{sec:highdim} The topological thermodynamics in higher dimensions}

It is interesting to determine the influence of the dimensional number on the topological thermodynamics of the regular black string reduced dimensionally on the circle. We can generalize the five-dimensional black string to the higher-dimensional case which is a solution of the ($D+1$)-dimensional Einstein-Maxwell theory (with $D>4$) compactified on the circle $S^1$ with respect to the double Wick rotation symmetry \cite{Ibrah2021}. The line element and ($D-2$)-form field strength are given by
\begin{eqnarray}
ds^{2}_{D+1}&=&-f_{S}(r)dt^{2}+f_{B}(r)dy^{2}+\frac{dr^{2}}{h(r)}+r^{2}d\Omega_{D-2}^{2},\nonumber\\
F&=&PdV_{S^{D-2}},\label{HDWDR-sol}
\end{eqnarray}
where 
\begin{eqnarray}
f_{B}(r)=1-\left(\frac{r_{b}}{r}\right)^{D-3},\quad f_{S}(r)=1-\left(\frac{r_{h}}{r}\right)^{D-3},\quad h(r)=f_{B}(r)f_{S}(r),   
\end{eqnarray}
$d\Omega_{D-2}^{2}$ and $dV_{S^{D-2}}$ refer to the line element and the volume form of unit ($D-2$)-sphere, respectively, and $P$ is the magnetic charge of the regular black string. The horizons of the regular black strings have a $S^{D-2}\times S^{1}$ topology. 

The dimensional reduction of the regular black strings on the circle leads to a solution of the $D$-dimensional Einstein-Maxwll-dilaton whose mass $M$ and magnetic charge $Q$ are given as follows
\begin{eqnarray}
M&=&\frac{\pi^{\frac{D-1}{2}}}{\Gamma\left(\frac{D-1}{2} \right) }\left[(D-2)r_{h}^{D-3}+r_{b}^{D-3} \right],\\
Q&=&\frac{(D-3)(D-1)r_{h}^{D-3}r_{b}^{D-3}}{2}. 
\end{eqnarray}
Note that, we have here set $\kappa_{D}=1$. The corresponding Bekenstein-Hawking entropy and the temperature are determined as follows
\begin{eqnarray}
S&=&\frac{4\pi^{\frac{D+1}{2}}}{\Gamma\left(\frac{D-1}{2} \right) }\left[r_{h}^{D-1}(r_{h}^{D-3}-r_{b}^{D-3}) \right]^{\frac{1}{2}},\\
T&=&\frac{D-3}{4\pi r_{h}} \sqrt{1-\left(\frac{r_{b}}{r_{h}} \right)^{D-3} }.
\end{eqnarray}

\begin{figure}[h]
  \includegraphics[scale=0.4]{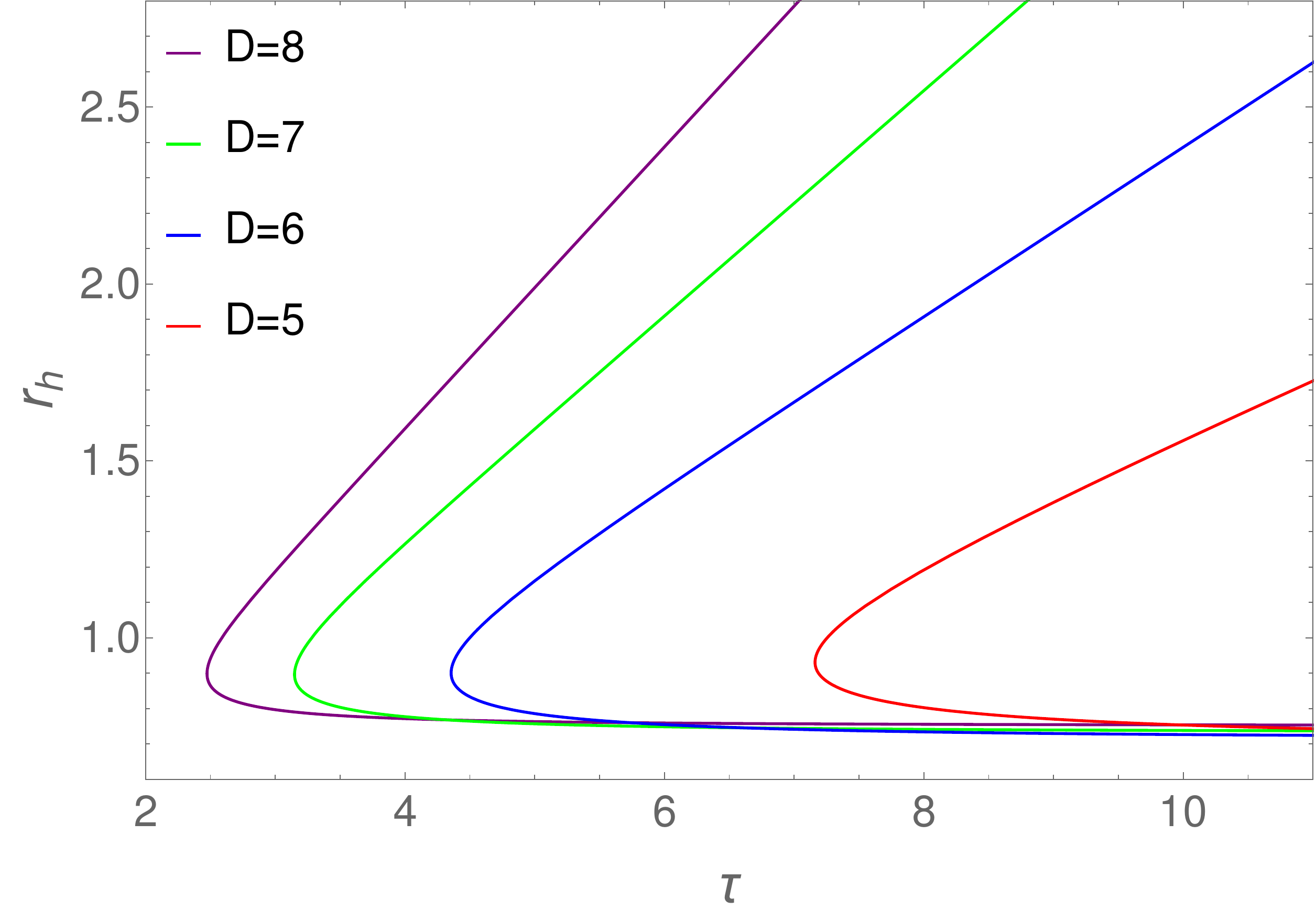}
  \caption{Locus of zero points shown in the $r_{h}-\tau$ plane for higher-dimensional black strings.}
  \label{fig:f7}
\end{figure}

\begin{figure}[t]
  \includegraphics[scale=0.4]{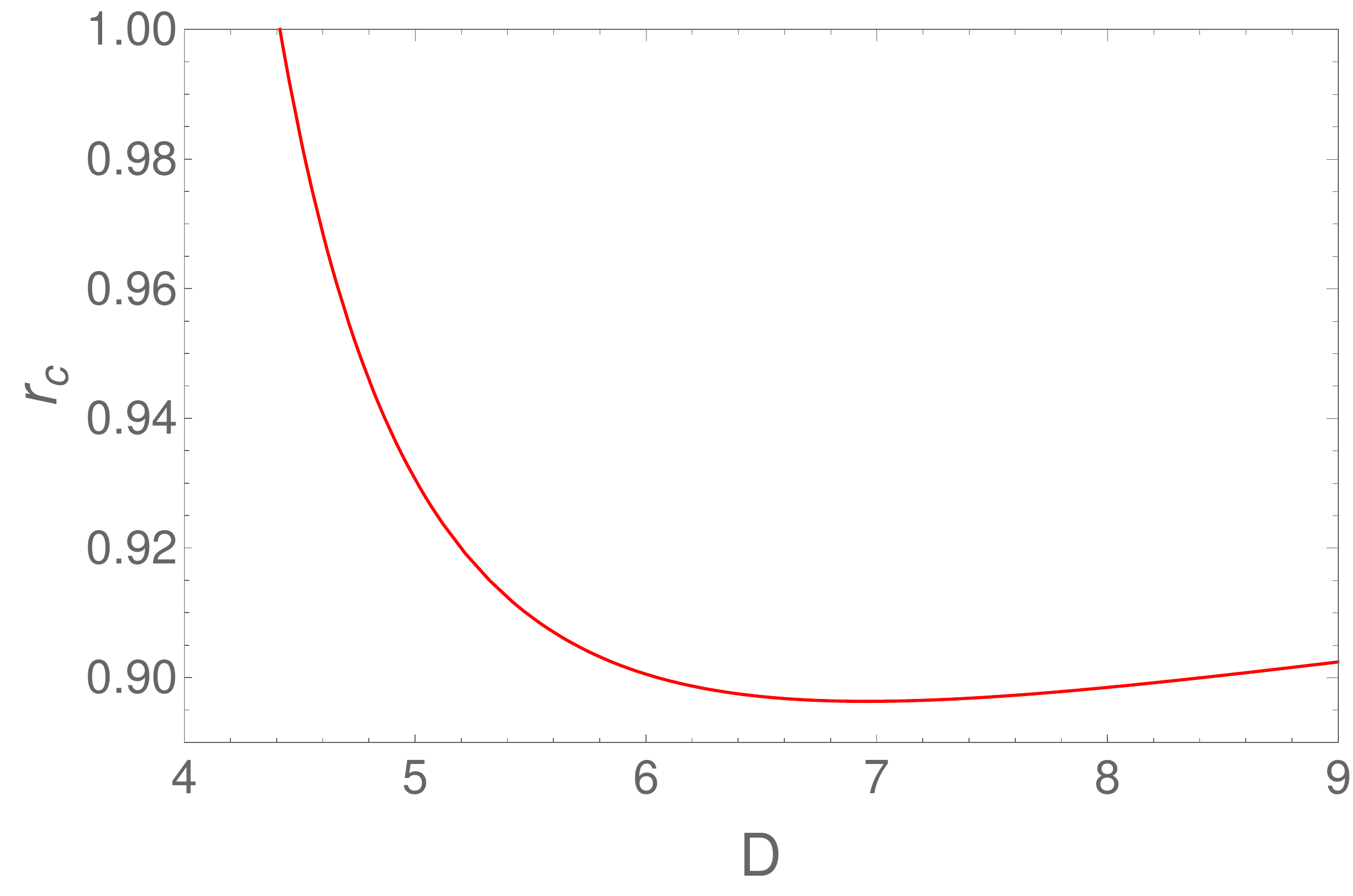}
  \caption{The critical horizon $r_c$ of the higher-dimensional regular black strings as a function of the dimensional number $D$.}
  \label{fig:f8}
\end{figure}

In order to study the topological thermodynamics of the above solution, we construct the generalized free energy as follows
\begin{eqnarray}
\mathcal{F}&=&M-\frac{S}{\tau}\nonumber\\
&=&\frac{\pi^{\frac{D-1}{2}}}{\Gamma\left( \frac{D-1}{2}\right)}\left[\frac{-4\pi}{\tau} \sqrt{r_{h}^{2D-4}-\frac{2r_{h}^{2}}{3-4D+D^2}} +\frac{2r_{h}^{3-D}}{3-4D+D^2}+(D-2)r_{h}^{D-3}\right] .
\end{eqnarray}
The zero points of the vector field $(\phi^{r_{h}},\phi^{\Theta})$ correspond to $\Theta=\pi/2$ and the values of $r_{h}$ satisfying the extremal condition $\frac{\partial \mathcal{F}}{\partial r_h}=0$ leading to the following equation
\begin{eqnarray}
4\pi r_{h}^{D}+(3-D)\left( r_{h}\sqrt{r_{h}^{2D-4}-\frac{2r_{h}^{2}}{3-4D+D^{2}}}\right) \tau=0.
\end{eqnarray}
By solving this equation, we can find the behavior of the event horizon in terms of $\tau$ at the zero points depending on the dimensional number. In Fig. \ref{fig:f7}, we show this behavior in the $(\tau,r_{h})$ plane for the cases of $D=5$, 6, 7, and 8. 
Similar to the five-dimensional black string, these curves exhibit two branches representing the small and large phases, respectively, and possess a distinctive phase transition point. It means that the thermodynamic topology of the higher-dimensional regular black strings is the same as the five-dimensional case, which is characterized by the local winding numbers of 1 and -1. The coordinates of the phase transition points are denoted by $(\tau_{c}, r_{c})$ determined by the following relations
\begin{eqnarray}
\tau_{c}&=&4\pi \sqrt{\frac{D-1}{D-3}}\frac{r_{c}^{D-1}}{\sqrt{(D-1)(D-3)r_{c}^{2D-4}-2r_{c}^{2}}},\\
r_{c}&=&\left(\frac{D^{2}-4D+3}{2D-4} \right) ^{\frac{1}{6-2D}}.
\end{eqnarray}
The behavior of $\tau_c$ in terms of the dimensional number can be seen in Fig. \ref{fig:f7}: $\tau_c$ monotonically decreases with the growth of the dimensional number. Whereas, the behavior of $r_c$ is depicted in Fig. \ref{fig:f8}. 
We observe that the critical radius behaves as a monotonically decreasing function of the dimensional number $D$ for the region of sufficiently small $D$ but it should monotonically increase with the growth of the dimensional number for the region of sufficiently large $D$. This point that changes this behavior of $r_c$ corresponds to $D=7$.

We can confirm the thermodynamically stable and unstable phases of the higher-dimensional regular black strings found by the topological method and the corresponding phase transition by investigating heat capacity at a fixed magnetic charge, given by
\begin{eqnarray}
C_{Q}&=&-\frac{4\pi^{\frac{1+D}{2}}}{\Gamma\left(\frac{D-1}{2} \right) }\frac{(D-1)(D-2)(D-3)r_{h}^{2D}-2Q^{2}r_{h}^{6}}{(D-1)(D-3)r_{h}^{2D}-2(D-2)Q^{2}r_{h}^{6}}r_{h}^{D-2}\sqrt{1-\frac{2Q^{2}r_{h}^{6-D}}{3-4D+D^{2}}}.
\end{eqnarray}
\begin{figure}[h]
  \includegraphics[scale=0.4]{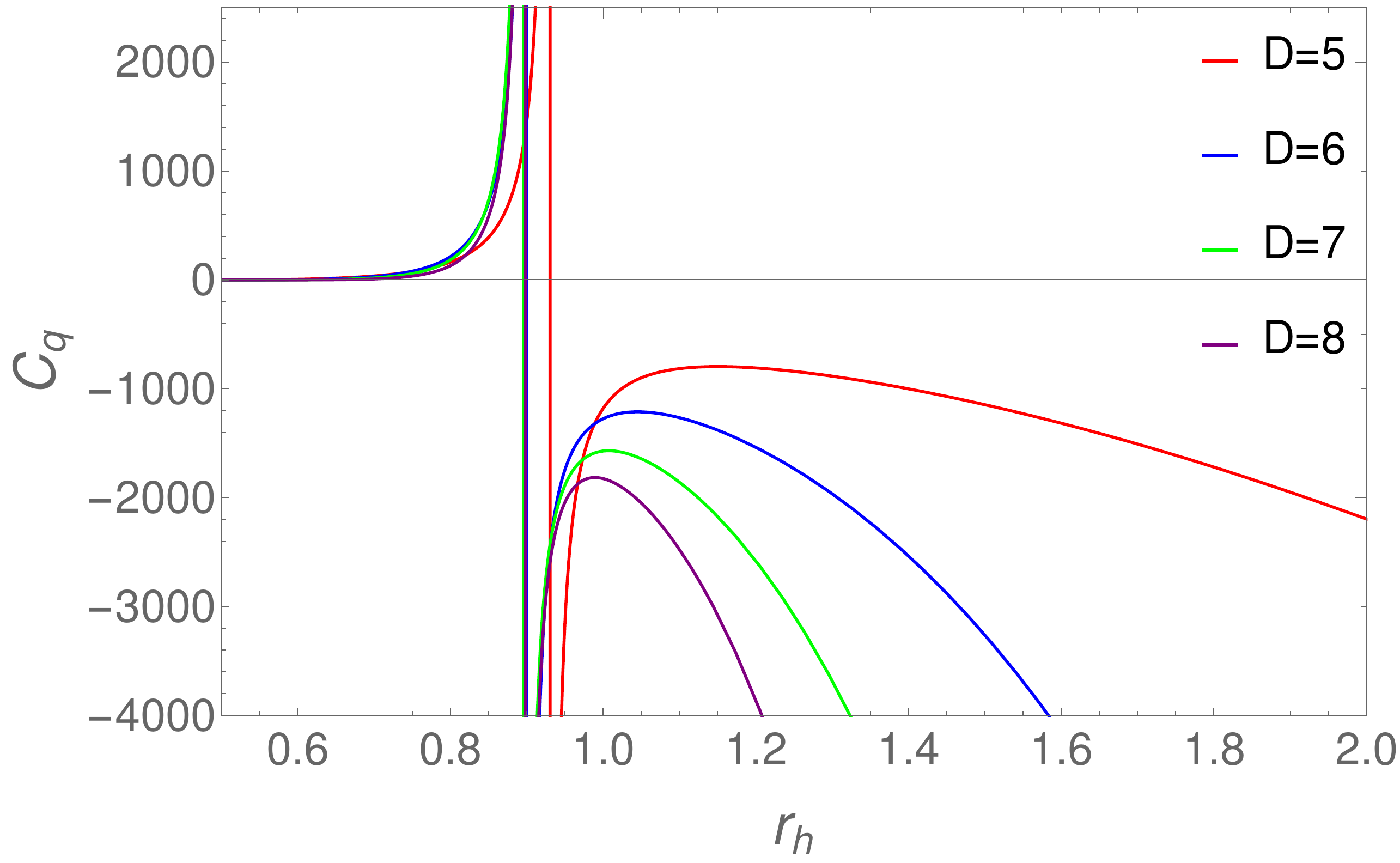}
  \caption{The heat capacity $C_{Q}$ as a function of $r_{h}$ for the higher-dimensional black strings with various values of the dimensional number.}
  \label{fig:f9}
\end{figure}
As shown in Fig. \ref{fig:f9}, the heat capacity $C_{Q}$ of the higher-dimensional regular black strings exhibits thermodynamically stable and unstable phases corresponding to the positive and negative values of $C_Q$. Furthermore, the heat capacity $C_{Q}$ is divergent at $r_c$ and as a result, there is a second-order phase transition between the higher-dimensional regular black strings that correspond to the local winding numbers of 1 and -1.

\section{\label{sec:conclu} Conclusion}

In this paper, we employ the topology method to investigate the general thermodynamic characteristics of regular black strings from the dimensional reduction perspective on a circle. Our results indicate that the dimensionally reduced regular black strings can be characterized by the local winding numbers that are 1 and -1 corresponding to the thermodynamic stability and the thermodynamic instability, respectively. The global winding number which is a sum of the local winding numbers is equal to zero in analogy to the Reissner-Nordström (RN) black hole. Furthermore, we confirm the topological classification for the phase structure of the dimensionally reduced regular black strings and its topological indication for the phase transition by studying the heat capacity at a fixed magnetic charge. The behavior of the heat capacity reveals that the zero point having a positive(negative) winding number corresponds to a positive(negative) heat capacity. This outcome validates the conjecture postulating that the zero point characterized by a positive(negative) winding number corresponds to a thermodynamic stable(unstable) black hole. Furthermore, the identification of the locus of topological defects in the parameter space defined by $(\tau,r_h)$ exhibits a generation point associated with a second-order phase transition corresponding to the divergence of the heat capacity.

In addition, in order to validate the results of the topological approach, we study the free energy of the dimensionally reduced regular black strings to find the thermodynamically stable and unstable phases and the phase transition between them. By examining the generalized free energy as a function of the horizon radius, we find two critical points which are the local minimum and maximum points of the generalized free energy for the case of $\tau^{-1}<\sqrt{3/128\pi^{2}}$. The local minimum(maximum) point of the generalized free energy corresponds to the thermodynamically stable(unstable) phase of the dimensionally reduced regular black strings. This result is completely compatible with the topological indication for the thermodynamic phases of the dimensionally reduced regular black strings. Also, we study the on-shell free energy in terms of the temperature, which exhibits a distinctive discontinuity at a specific temperature value $T=\sqrt{3/128\pi^{2}}$. This abrupt change in the on-shell free energy suggests the presence of a second-order phase transition between the small and large regular black strings.

The topological approach is also confirmed by the investigation of the Ruppeiner scalar curvature on the geometry of the state space of the dimensionally reduced regular black strings. The Ruppeiner scalar curvature is divergent at $r_{c}=\sqrt{4/3}$ where the heat capacity is too. Hence, it indicates a thermodynamic instability of the dimensionally reduced regular black string corresponding to a phase transition between the thermodynamically stable and unstable phases. The fact that the Ruppeiner scalar curvature of the dimensionally reduced regular black strings is always negative suggests that the interactions among their microstructures are only attractive.


\begin{thebibliography}{99}
\bibitem{Hawk1975} S. W. Hawking, Commun. Math. Phys. {\bf 43}, 199 (1975).

\bibitem{Bekenstein1972} J. D. Bekenstein, Lett. Nuovo Cim. {\bf 4}, 737 (1972).

\bibitem{Bardeen1973} J. M. Bardeen, B. Carter, and S. W. Hawking, Commun. Math. Phys. {\bf 31}, 161-170 (1973)

\bibitem{Hawk1976} S. W. Hawking, Phys. Rev. D {\bf 13}, 191 (1976).

\bibitem{SWang2006} S. Wang, S.-Q. Wu, F. Xie, and L. Dan, Chin. Phys. Lett. {\bf 23}, 1096 (2006).

\bibitem{Kastor2009} D. Kastor, S. Ray, and J. Traschen, Class. Quantum Grav. {\bf 26}, 195011 (2009).

\bibitem{Kastor2010} D. Kastor, S. Ray, and J. Traschen, Class. Quantum Grav. {\bf 27}, 235014 (2010).

\bibitem{Dolan2011} B. P. Dolan, Class. Quantum Grav. {\bf 28}, 125020 (2011).

\bibitem{Kubiznak2017} D. Kubiznak, R. B. Mann, and M. Teo, Class. Quantum Grav. {\bf 34}, 063001 (2017).



\bibitem{Kubiznak2012} D. Kubiznak and R. B. Mann, JHEP {\bf 1207}, 033 (2012).

\bibitem{Hendi2013} S. H. Hendi and M. H. Vahidinia, Phys. Rev. D {\bf 88}, 084045 (2013).

\bibitem{Cai2013} R.-G. Cai, L.-M. Cao, L. Li, and R.-Q. Yang, JHEP {\bf 1309}, 005 (2013).

\bibitem{XMo2014} J.-X. Mo and W.-B. Liu, Eur. Phys. J. C {\bf 74}, 2836 (2014).

\bibitem{Hendi2016} S. H. Hendi, R. M. Tad, Z. Armanfard, and M. S. Talezadeh, Eur. Phys. J. C {\bf 76}, 263 (2016).

\bibitem{Nam2018b} C. H. Nam, Eur. Phys. J. C {\bf 78}, 1016 (2018).

\bibitem{Nam2019} C. H. Nam, Gen. Rel. Grav. {\bf 51}, 100 (2019).



\bibitem{Johnson2014} C. V. Johnson, Class. Quantum Grav. {\bf 31}, 205002 (2014).

\bibitem{Setare2015} M. R. Setare and H. Adami, Gen. Rel. Grav. {\bf 47}, 133 (2015).

\bibitem{Bhamidipati2017}  C. Bhamidipati and P. K. Yerra, Eur. Phys. J. C {\bf 77}, 534 (2017).

\bibitem{XMo2018} J.-X. Mo and S.-Q. Lan, Eur. Phys. J. C {\bf 78}, 666 (2018).

\bibitem{Santos2018} J. F. G. Santos, Eur. Phys. J. Plus {\bf 133}, 321 (2018).

\bibitem{Zhang2019}  J. Zhang, Y. Li, and H. Yu, JHEP {\bf 1902}, 144 (2019).

\bibitem{Nam2021a} C. H. Nam, Gen. Rel. Grav. {\bf 53}, 30 (2021).



\bibitem{Aydner2017a} \"{O}. \"{O}kc\"{u} and E. Aydner, Eur. Phys. J. C {\bf 77}, 24 (2017).

\bibitem{Aydner2017b} \"{O}. \"{O}kc\"{u} and E. Aydner, Eur. Phys. J. C {\bf 78}, 123 (2018).

\bibitem{Mo-Li2018} J.-X. Mo, G.-Q. Li, S.-Q. Lan, and X.-B. Xu, Phys. Rev. D {\bf 98}, 124032 (2018). 

\bibitem{QLan2018} S.-Q. Lan, Phys. Rev. D {\bf 98}, 084014 (2018).

\bibitem{Nam2020} C. H. Nam, Eur. Phys. J. Plus {\bf 135}, 259 (2020).


\bibitem{Hawking1973} S. W. Hawking and G. F. R. Ellis, \emph{The Large Scale Structure of Spacetime} (Cambridge University Press, Cambridge, 1973).



\bibitem{Ayon-Beato98} E. Ay\'{o}n-Beato and A. Garc\'{i}a, Phys. Rev. Lett. {\bf 80}, 5056 (1998).

\bibitem{Ayon-Beato2000} E. Ay\'{o}n-Beato and A. Garc\'{i}a, Phys. Lett. B {\bf 493}, 149 (2000).

\bibitem{Bronnikov2001} K. A. Bronnikov, Phys. Rev. D {\bf 63}, 044005 (2001).

\bibitem{Berej2006} W. Berej, J. Matyjasek, D. Tryniecki, and M. Woronowicz, Gen. Rel. Grav. {\bf 38}, 885 (2006).

\bibitem{Gonzalez2009} H. A. Gonzalez and M. Hassaine, Phys. Rev. D {\bf 80}, 104008 (2009).

\bibitem{Toshmatov2014} B. Toshmatov, B. Ahmedov, A. Abdujabbarov, Z. Stuchlik, Phys. Rev. D {\bf 89}, 104017 (2014).

\bibitem{Ghosh2015} S. G. Ghosh and S. D. Maharaj, Eur. Phys. J. C {\bf 75}, 7 (2015).

\bibitem{Hendi2015} S. H. Hendi and  A. Dehghani, Phys. Rev. D {\bf 91}, 064045 (2015).

\bibitem{Dehghani2017} M. Dehghani and S. F. Hamidi, Phys. Rev. D {\bf 96}, 044025 (2017).

\bibitem{Rincon2107} A. Rinc\'{o}n, B. Koch, P. Bargue\~{n}o, G. Panotopoulos, and A. H. Arboleda, Eur. Phys. J. C {\bf 77}, 494 (2017).

\bibitem{Nam2018a} C. H. Nam, Eur. Phys. J. C {\bf 78}, 418 (2018).

\bibitem{Nam2018c} C. H. Nam, Eur. Phys. J. C {\bf 78}, 581 (2018).

\bibitem{Singh2018} S. G. Ghosh, D. V. Singh, and S. D. Maharaj, Phys. Rev. D {\bf 97}, 104050 (2018).

\bibitem{Hyun2019} S. Hyun and C. H. Nam, Eur. Phys. J. C {\bf 79}, 737 (2019).



\bibitem{Ibrah2021} I. Bah and P. Heidmann, Phys. Rev. Lett. {\bf 126}, 151101 (2021).

\bibitem{Nam2023} C. H. Nam, \emph{Microstates and statistical entropy of observed black holes}, arXiv:2304.04491.

\bibitem{Nam2023c} C. H. Nam, Phys. Lett. B {\bf 841}, 137930 (2023).

\bibitem{Arkani-Hamed1998} N. Arkani-Hamed, S. Dimopoulos, and G. R. Dvali, Phys. Lett. B {\bf 429}, 263 (1998).

\bibitem{Randall1999} L. Randall and R. Sundrum, Phys. Rev. Lett. {\bf 83}, 3370 (1999).

\bibitem{Nam2021} C. H. Nam, Eur. Phys. J. C {\bf 81}, 1102 (2021).

\bibitem{Nam2023b} C. H. Nam, Phys. Rev. D {\bf 107}, 063502 (2023). 

\bibitem{Nam2023a} C. H. Nam, Phys. Rev. D {\bf 107}, L041901 (2023).

\bibitem{Miyamoto2006} U. Miyamoto and H. Kudoh, JHEP {\bf 12} (2006) 048.

\bibitem{Stotyn2011} S. Stotyn and R. B. Mann, Phys. Lett. B {\bf 705}, 269 (2011).

\bibitem{Hung2023} T. N. Hung and C. H. Nam, \emph{Compactified extra dimension and entanglement island as clues to quantum gravity}, arXiv:2303.00348.

\bibitem{RanLi2020} R. Li and J. Wang, Phys. Rev. D {\bf 102}, 024085 (2020).

\bibitem{York1986} J. W. York, Phys. Rev. D {\bf 33}, 2092 (1986).


\bibitem{Andr2020} R. André and J. P. S. Lemos, Phys. Rev. D {\bf 102}, 024006 (2020).

\bibitem{RanLi22020} R. Li, K. Zhang, and J. Wang, JHEP {\bf 2020} (2020) 90.

\bibitem{SWWei2022} S.-W. Wei, Y.-X. Liu, and R. B. Mann, Phys. Rev. Lett. {\bf 129}, 191101 (2022).

\bibitem{Cunha2020} P. V. P. Cunha and C. A. R. Herdeiro, Phys. Rev. Lett. {\bf 124}, 181101 (2020).

\bibitem{Duan1979} Y. S. Duan and M. L. Ge, Sci. Sin. {\bf 9}, 1072 (1979).

\bibitem{LBFu2000} L. B. Fu, Y. S. Duan, and H. Zhang, Phys. Rev. D {\bf 61}, 045004 (2000). 

\bibitem{FangZhang2023} C. Fang, J. Jiang, and M. Zhang, JHEP {\bf 01} (2023) 102.


\bibitem{Yerra2022} P. K. Yerra, C. Bhamidipati, and S. Mukherji, Phys. Rev. D {\bf 106}, 064059 (2022).

\bibitem{CLiu2023} C. Liu and J. Wang, Phys. Rev. D {\bf 107}, 064023 (2023).

\bibitem{Liwang2023} R. Li and J. Wang, \emph{Generalized free energy landscapes of the charged Gauss-Bonnet AdS black holes in diverse dimensions}, arXiv:2304.03425.

\bibitem{NCBai2023} N. C. Bai, L. Li, and J. Tao, Phys. Rev. D {\bf 107}, 064015 (2023).

\bibitem{DiWu2023} D. Wu, Phys. Rev. D {\bf 107}, 024024 (2023).

\bibitem{DiWu2023-2} D. Wu and S. Q. Wu, Phys. Rev. D {\bf 107}, 084002 (2023).

\bibitem{Wein1975}  F. Weinhold, J. Chem. Phys. {\bf 63}, 2479 (1975).

\bibitem{Ruppeiner1983} G. Ruppeiner, Phys. Rev. Lett. {\bf 50}, 287 (1983).

\bibitem{Rupp1995} G. Ruppeiner, Rev. Mod. Phys. {\bf 67}, 605 (1995).

\bibitem{Mrug1984} R. Mrugala, Physica {\bf 125A}, 631 (1984).

\bibitem{Salamon1984} P. Salamon, J. Nulton, and E. Ihrig, J. Chem. Phys. {\bf 80}, 436 (1984).

\bibitem{Jany1990} H. Janyszek and R. Mrugala, J. Phys. A: Math. Gen. {\bf 23}, 467 (1990).

\end{thebibliography}
\end{document}